\documentclass[a4paper]{jpconf}
\usepackage{graphicx}
\usepackage{lineno}
\usepackage{url}
\begin{document}
\title{Heavy Flavor Physics in PHENIX}

\author{Rachid Nouicer, for the PHENIX Collaboration
}

\address{Physics Department, Brookhaven National Laboratory, Upton, NY 11973, USA.}

\ead{rachid.nouicer@bnl.gov}

\begin{abstract}
Heavy quarks are good probes of the hot and dense medium created in
relativistic heavy ion collisions since they are mainly generated
early in the collision and interact with the medium in all collision
stages. In addition, heavy flavor quarkonia production is thought to
be uniquely sensitive to the deconfined medium of the Quark Gluon
Plasma (QGP) through color screening.  Heavy quark production has been
studied by the PHENIX experiment at RHIC via measurements of single
leptons from semi-leptonic decays, in both the electron channel at
mid-rapidity and in the muon channel at forward rapidity. Large
suppression and azimuthal anisotropy of single electrons have been
observed in Au+Au collisions at ${\rm \sqrt{s_{NN}}}$
=\ 200\,GeV. These results suggest a large energy loss and strong flow
of the heavy quarks in the hot, dense matter. The PHENIX experiment
has also measured J/$\psi$ production in $p+p$, d+Au, Cu+Cu, and Au+Au
collisions at energies up to ${\rm \sqrt{s_{NN}}}$ =\ 200\,GeV. In 
central Au+Au at 200 GeV, more suppression is observed at forward
rapidity than at central rapidity. This can be interpreted either as a
sign of quark recombination, or as a hint of additional cold nuclear
matter effects. Selected PHENIX results on open heavy flavor and heavy
quarkonia production in $p+p$, d+Au, Cu+Cu, and Au+Au collisions are
presented.  The status of the recent PHENIX upgrade of the central
Silicon Vertex Tracker (VTX) and its performance are elucidated.
\end{abstract}

\section{Introduction}

The measurement of inclusive hadron yields in central Au+Au collisions
at RHIC led to the discovery of the suppression of hadron production
at large transverse momenta ($p_T$) compared to $p+p$
collisions~\cite{RHIC}. This is generally attributed to the energy
loss of light partons in the dense nuclear matter created at RHIC.
Heavy quarks, i.e. charm and beauty, are believed to be mostly created
in initial hard scattering processes of partons~\cite{Lin95} and thus
are excellent probes of the hot and dense matter formed in
nucleus$-$nucleus collisions at high energy. While some of the
produced pairs form bound quarkonia, the vast majority hadronize into
hadrons carrying open heavy flavor. They interact with the medium and
are expected to be sensitive to its energy density through the
mechanism of parton energy loss. Due to the large mass of heavy
quarks the suppression of small angle gluon radiation should reduce
their energy loss, and consequently any suppression of heavy-quark
mesons like $D$ and $B$ mesons at high $p_T$ is expected to
be smaller than that observed for hadrons consisting of light
quarks~\cite{Dok01}. 

Heavy quark production has been studied by the PHENIX experiment at
RHIC via measurements of single leptons from semi-leptonic decays in
both the electron channel at mid-rapidity and in the muon channel at
forward rapidity. In this paper I will summarize the latest PHENIX
results concerning open and closed heavy flavor production as a
function of beam energy and systems size, and will give an overview of the
latest progress with the Silicon Vertex Tracker (VTX).
\section{PHENIX Experiment}

The PHENIX detector~\cite{PHE03} comprises three separate spectrometers in
three pseudorapidity ($\eta$) ranges. Two central arms at midrapidity
cover $\mid \eta \mid <$ 0.35 and have an azimuthal coverage ($\phi$)
of $\pi$/2 rad each, while muon arms at backward and forward rapidity
cover -2.2 $< \eta <$ -1.2 and 1.2 $< \eta <$ 2.4, respectively, with
full azimuthal coverage. In the central arms, heavy quark
production was studied via measurements of single leptons (electrons) from
semi-leptonic decays. Charged particle tracks are reconstructed using
the drift chamber and pad chambers. Electron candidates are selected
by matching charged tracks to hits in the Ring Imaging Cherenkov (RICH)
counters and clusters in the Electromagnetic Calorimeter (EMCal). At
forward and backward rapidity, the heavy quark production is measured
via dimuon decays. Muons are identified by matching tracks measured in
cathode-strip chambers, referred to as the muon tracker (MuTr), to
hits in alternating planes of Iarocci tubes and steel absorbers,
referred to as the muon identifier (MuID). Each muon arm is located
behind a thick copper and iron absorber that is meant to stop most
hadrons produced during the collisions, so that the detected muons
must penetrate 8 to 11 hadronic interaction lengths of material in total. Beam
interactions are selected with a minimum-bias (MB) trigger requiring
at least one hit in each of the two beam-beam counters (BBCs) located at
positive and negative pseudorapidity 3~$< \mid \eta \mid <$~3.9.
\begin{figure}[!]                                                                                    
\begin{minipage}{18.5pc}                                                                             
\includegraphics[width=18pc]{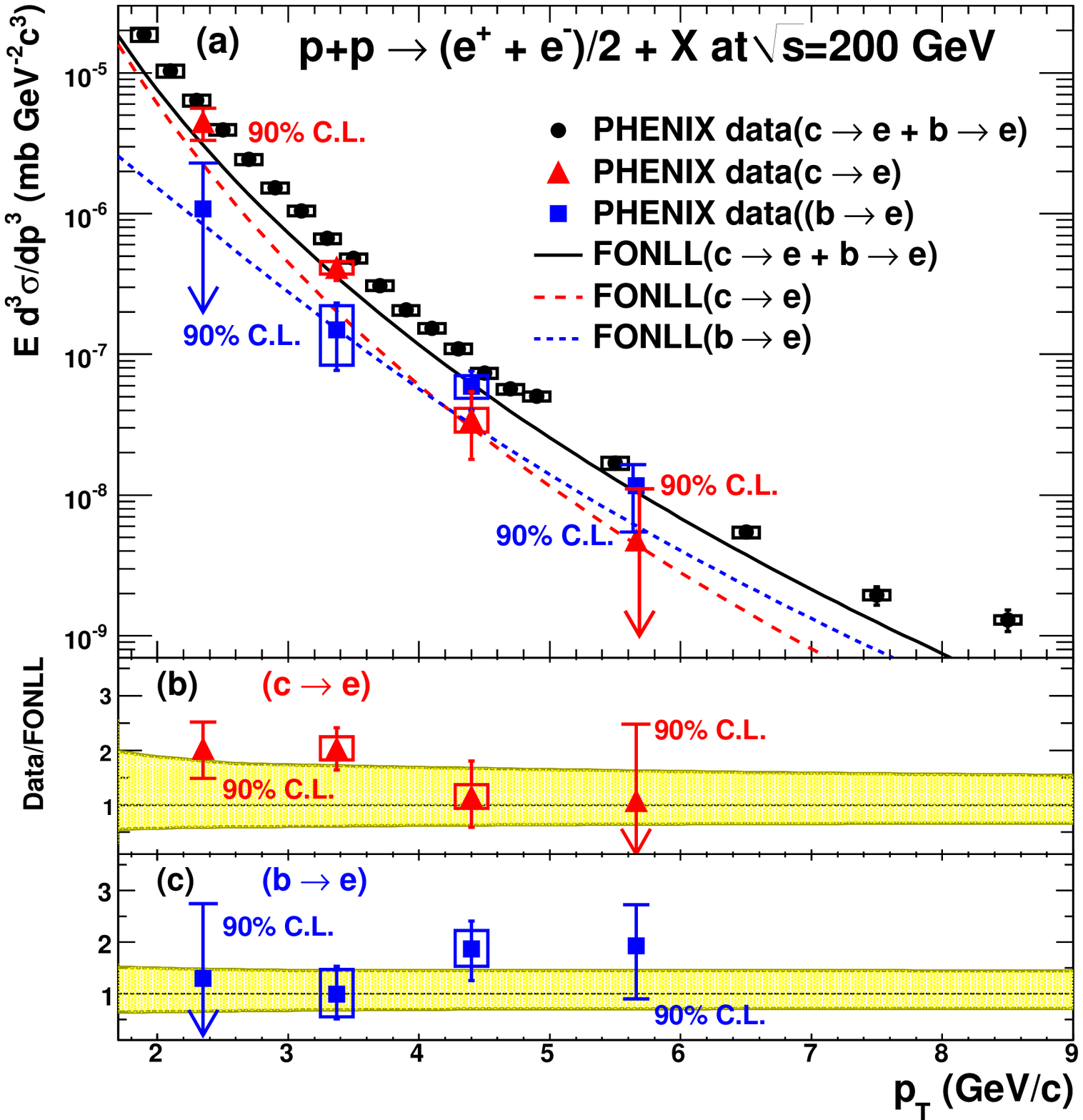}                                                             
\caption{\small \label{fig1} Invariant cross-sections of electrons
  from charm and bottom with the FONLL calculation in $p+p$ collisions
  at ${\rm \sqrt{s_{_{NN}}}}$ =\ 200\,GeV. Panels (b) and (c) show the
  ratios of data points over the FONLL prediction as a function of
  electron $p_T$ for charm and bottom, respectively. The shaded area
  shows the uncertainty in the FONLL prediction~\cite{PHE06,PHE09}.}
\end{minipage}
\hspace{0.5pc}%
\begin{minipage}{18pc}                                                                               
\includegraphics[width=18pc]{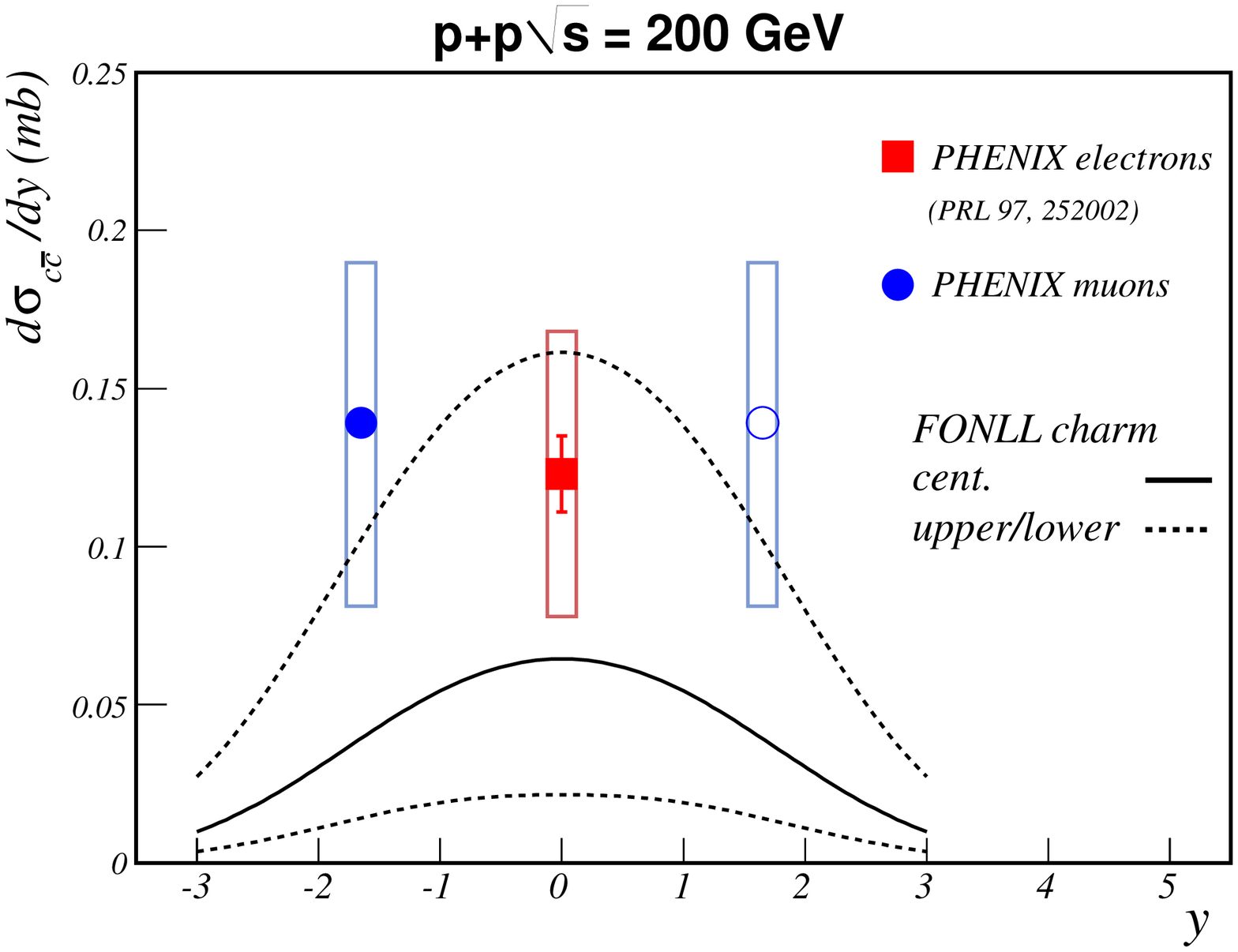}
\caption{\small \label{fig2} Heavy flavor, $(c\bar{c}$),
  production cross-section as a function of rapidity in $p+p$
  collisions at ${\rm \sqrt{s_{_{NN}}}}$ =\ 200\,GeV, measured using
  semi-leptonic decay to electrons (closed square) and to muons
  (closed circles)~\cite{PHE12A}.}
\end{minipage}
\end{figure}        
\section{Open heavy flavor}
Open heavy flavor production is measured in PHENIX through
the measurement of inclusive electrons or muons. For the electron
measurement, the electrons that come from either meson decay or photon
conversions are measured and subtracted from the inclusive spectrum
and the remainder is attributed to electrons coming from the
semi-leptonic decay of $D$~and~$B$ mesons. This remainder is also
referred to as the non-photonic electron component. PHENIX has
measured spectra of the single electrons~\cite{PHE06,PHE09} and single
muons~\cite{PHE07} from heavy flavor in $p+p$ collisions at 200\,GeV as
well single electrons from heavy flavor in
Au$+$Au~\cite{PHEAuAu,RHICNouicer}, Cu$+$Cu~\cite{PHE12A,PHECuCu} and
d$+$Au~\cite{PHEdAu} collisions. Figure~\ref{fig1}(a) shows the measured
single electron spectra from charm (triangles) and bottom (squares)
compared to the FONLL predictions~\cite{PHE09,Cac05}. The measured
spectrum of single electrons (circles) is also shown for reference.
The single electron from charm and bottom are measured from the ratio,
(b $\rightarrow$ e)/(c $\rightarrow$ e + b $\rightarrow$ e), and the
spectrum of the electrons are from heavy flavor decays. Panel (b) ((c))
shows the ratio of the measured charm (bottom) cross-sections to the FONLL
calculation for charm production. 
The shaded area shows the uncertainty in the FONLL
prediction. The larger mass makes this uncertainty smaller in the case
of bottom quarks. These calculations agree with the data for bottom
production. The same is true for charm within the theoretical
uncertainty.

\begin{figure}[h]
\begin{minipage}{18pc}                                                                               
\includegraphics[width=18pc]{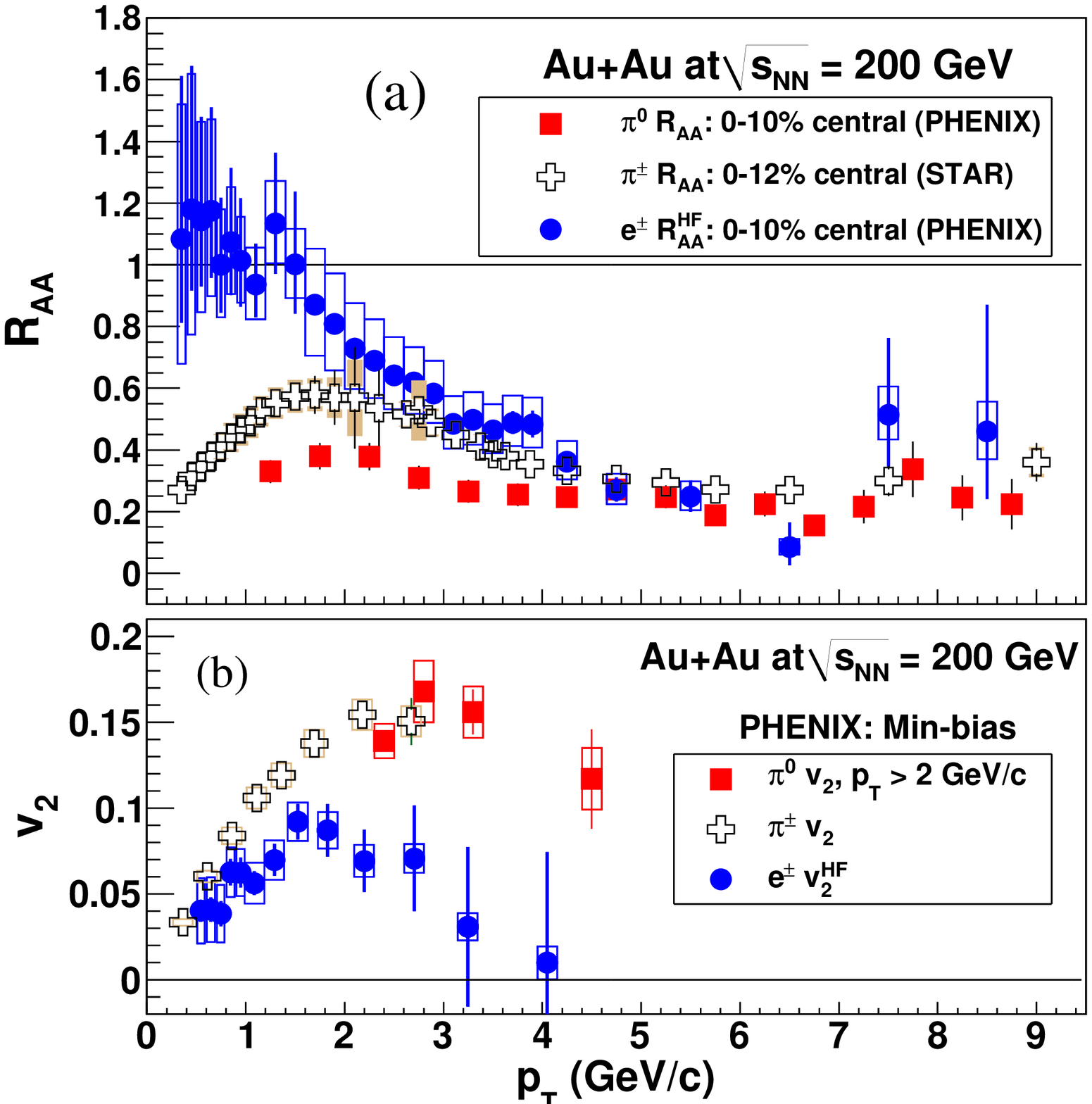}
\caption{\small \label{fig3} Nuclear modification factor, ${\rm
    R_{AA}^{HF}}$, for HF electrons compared with the ${\rm R_{AA}}$
  of $\pi^{0}$ in central Au+Au\ collisions at ${\rm
    \sqrt{s_{_{NN}}}}$ =\ 200\,GeV, see panel (a). Panel (b) considers
  the anisotropic flow of HF electrons v$_{2}^{HF}$ with that of
  v$_{2}$ of $\pi^{0}$ and $\pi^{\pm}$ in minimum-bias
  Au+Au\ collisions~\cite{PHEAuAu,RHICNouicer}.} 
\end{minipage}                                                                                       
\hspace{0.5pc}%
\begin{minipage}{18pc}                                                                             
\includegraphics[width=18pc]{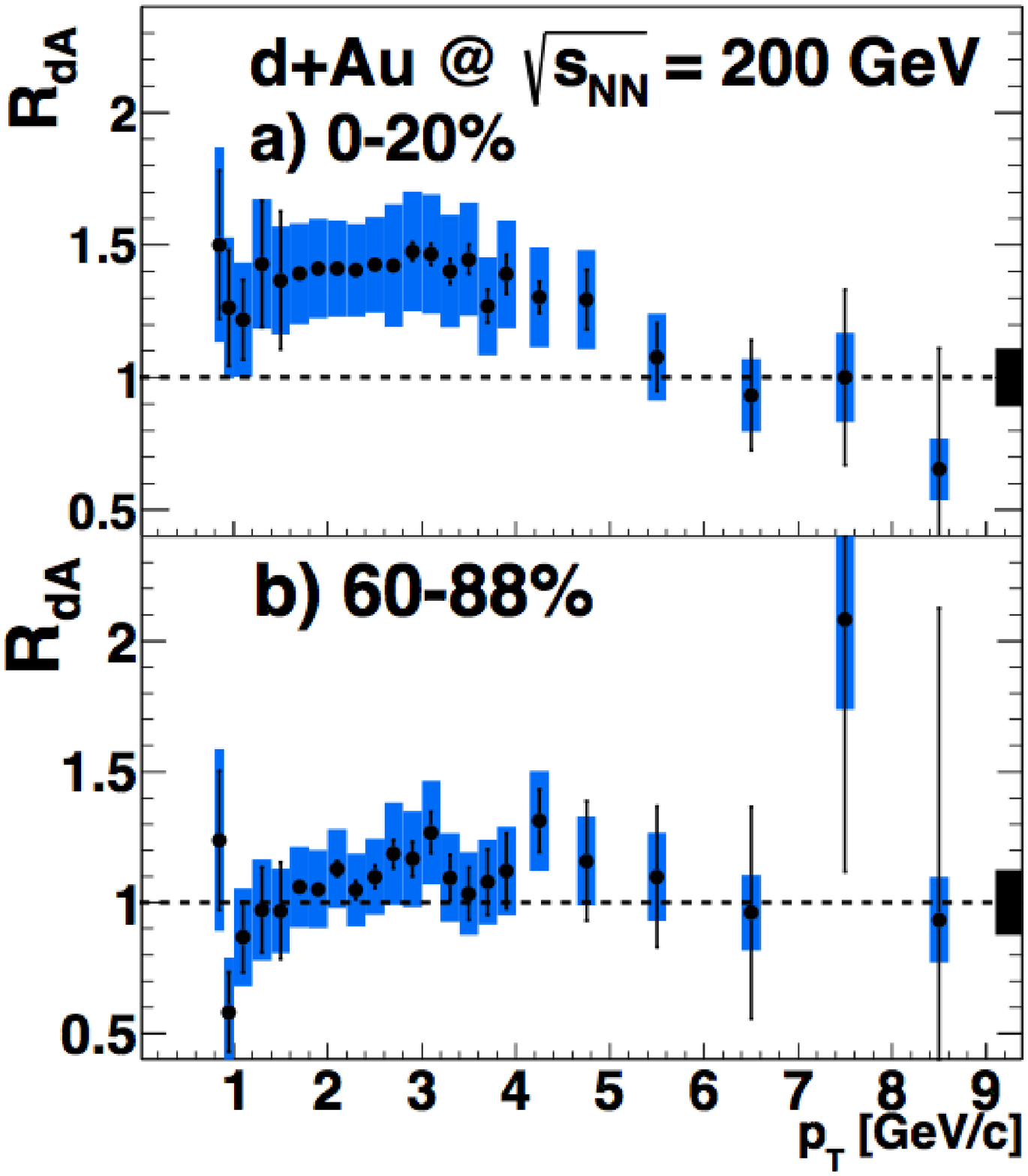}                                               
\caption{\small \label{fig4} Nuclear modification factor, ${\rm
    R_{dAu}^{HF}}$, for HF electrons measured in d+Au collisions at
  ${\sqrt{s_{_{NN}}}}$ =\ 200\,GeV for a) the most central collisions and
for b) the most peripheral collisions~\cite{PHEdAu}.}
\end{minipage}
\end{figure}

Heavy-flavor spectra in $p+p$ collisions, are measured via single leptons from
semi-leptonic decays in both the electron channel at mid-rapidity and
in the muon channel at forward rapidity as a function of $p_T$.
These spectra are used to estimate the charm differential production
cross-section, d$\sigma_{c\bar{c}}$/dy \cite{PHE12A}. Estimation of
the full charm cross-section requires a theoretical calculation in
order to extrapolate the measurement down to $p_T$\,=\,0\ GeV/$c$.
A set of fixed-order-plus-next-to-leading-log (FONLL)
calculations as shown on Fig.~\ref{fig1} have been
used~\cite{PHE12A}. Additionally, the contribution of bottom quark decay to
the heavy flavor electron (muon) $p_T$ distribution becomes
increasingly relevant for $p_T$$>$4\,GeV/$c$, but has a
negligible contribution to the integral and is ignored hereafter.
Using this method, the integrated charm production cross-section at
mid- and forward-rapidity are shown in Fig~\ref{fig2}.

In heavy ion collisions, the RHIC experiments have revealed a suppression
of the high transverse momentum component of hadron spectra at
mid-rapidity in central Au+Au collisions compared to scaled momentum
spectra from $p+p$ collisions at the same energy, ${\rm
  \sqrt{s_{NN}}}$ =\ 200\,GeV as shown in Fig.~\ref{fig3}a. This
effect, originally proposed by Bjorken, Gyulassy, and others
\cite{Bjorken} rests on the expectation of a large energy loss of high
momentum partons scattered in the initial stages of collisions in a
medium with a high density of free color charges. According to QCD
theory, colored objects may lose energy by the bremsstrahlung
radiation of gluons \cite{Gaard}. Such a mechanism would strongly
degrade the energy of leading partons, reflected in the reduced
transverse momentum of leading particles in the jets emerging after
fragmentation into hadrons. Adding to this discovery of suppression of
particles at high transverse momentum, two very striking results have
been seen for open heavy flavor from the PHENIX experiment via the
measurement of electrons from semi-leptonic decays of hadrons carrying
charm or bottom quarks. First, heavy mesons, despite their large mass,
exhibit a suppression at high transverse momentum compared to that
expected from $p+p$ interactions. This suppression is found to be
similar to that of light mesons, which implies a substantial energy
loss of fast heavy quarks while traversing the medium, see
Fig.~\ref{fig3}a; this shows the nuclear modification factors
measured for different types of particle in Au+Au collisions at 200
GeV.  The nuclear modification factor is defined as:
$${\rm R_{AA} (p_{T})= \frac{yield\ per\ A+A\ collisions}{N_{bin}
    \times(yield\ per\ p+p\ collisions) }}$$
$${\rm = \frac{d^{2}N^{^{A+A}}/dp_{T}d\eta}{N_{bin}\
  d^{2}N^{^{p+p}}/dp_{T}d\eta}}$$  
It involves scaling measured distributions of nucleon-nucleon transverse 
momentum by the number of expected incoherent binary collisions, 
${N_{bin}}$ \cite{RAAdefinition}.
Secondly, an elliptic
flow is observed for heavy mesons which is comparable to that of light
mesons like pions, see Fig.~\ref{fig3}b. This implies that the 
heavy quarks are in fact sensitive to the pressure gradients driving
hydrodynamic flow giving new insights into the strongly coupled nature
of the QGP fluid at these temperatures. 

\begin{figure}[h]
\begin{center}                                                                                    
\begin{minipage}{18pc}                                                                               
\includegraphics[width=20pc]{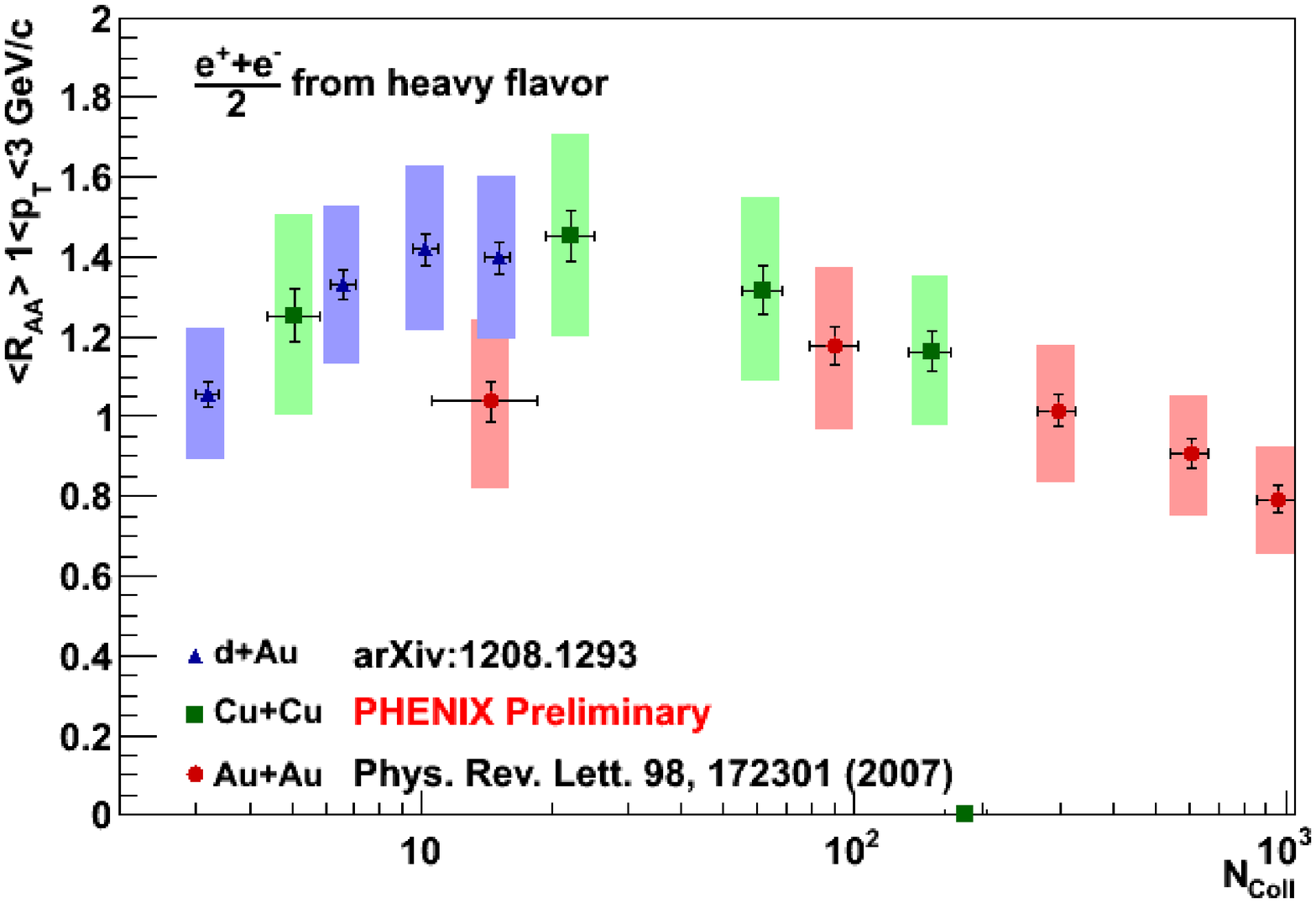}
\caption{\small \label{fig5} Nuclear modification factor for HF
  electrons for Cu+Cu collisions compared with those for d+Au and
  Au+Au. The ${\rm R_{A(d)A}^{HF}}$ are measured at mid-rapidity and
  integrated over the range $p_T$ = {1--3}~GeV/$c$.}
\end{minipage} 
\hspace{0.5pc}%
\begin{minipage}{18pc}                                                                               
\includegraphics[width=20pc]{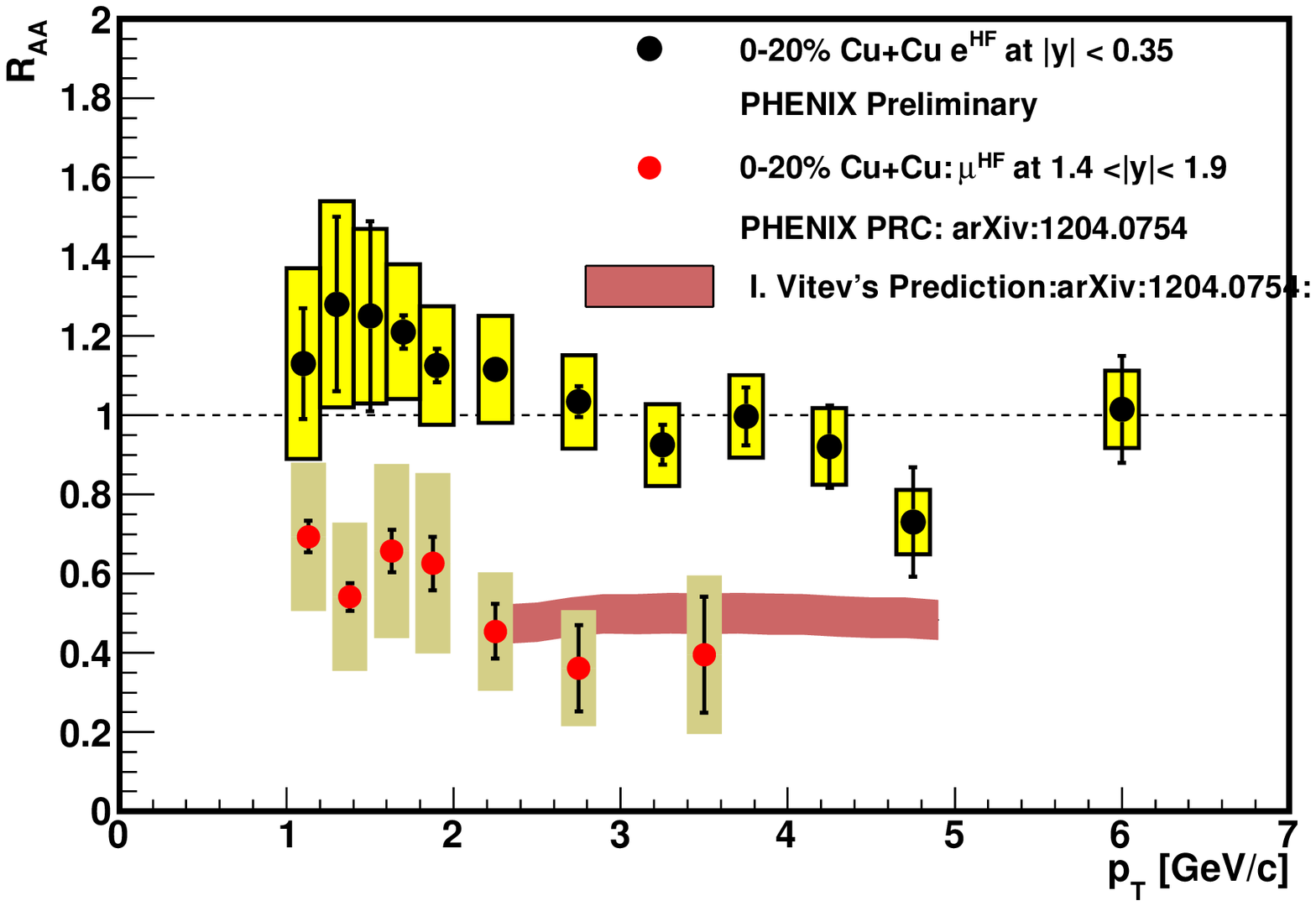}
\caption{\small \label{fig6} Nuclear modification factor ${\rm
    R_{CuCu}^{HF}}$ for semileptonic decays of heavy quarks in Cu+Cu
  collisions at 200\,GeV. The data at forward rapidity is compared to
  the theory curve~\cite{PHE12A,Sharma}.}
\end{minipage}
\end{center}                                                                                       
\end{figure}        

Recently, PHENIX has measured new results on open heavy flavor (HF)
production in d+Au collisions via semi-leptonic decays at mid-rapidity
[2]. The $p_T$ dependence of the nuclear modification is shown in
Fig~\ref{fig4} for central and peripheral collisions. Enhancement is
observed for the 0-20\% most central collisions in the $p_T$ range
1--5\,GeV/$c$, while little or no enhancement is observed for
peripheral collisions. At high $p_T$, where strong
suppression is observed for HF electrons in Au+Au collisions, the
measured ${\rm R_{dAu}^{HF}}$ is consistent with unity. Thus, the
suppression observed in Au+Au collisions can be attributed, within the
uncertainties on ${\rm R_{dAu}^{HF}}$, to hot nuclear matter
effects. There are new semi-leptonic decay open heavy flavor PHENIX
results for heavy ions that include Cu+Cu ${\rm R_{AA}^{HF}}$ data at
both mid- and forward-rapidity. In Fig.~\ref{fig5}, the mid-rapidity
Cu+Cu ${\rm R_{AA}^{HF}}$ data are compared as a function of
$N_{coll}$ with data from d+Au and Au+Au collisions. The data are
shown for $p_T$ range 1 to 3 GeV/$c$.  Where they overlap, the three
data sets are found to display similar behavior with $N_{coll}$ within
uncertainties. It is noteworthy that the d+Au data and the peripheral
Cu+Cu data have very similar modifications (including some enhancement
at $N_{coll}$ $\sim$ 10).  However, a comparison of forward (muon) and
midrapidity (electron) ${\rm R_{AA}^{HF}}$ in central (0--20\%) Cu+Cu
collisions shown an unexpected discrepancy~\cite{PHE12A,Sharma}.  We
observe that ${\rm R_{AA}^{HF}}$ (Cu+Cu) at forward rapidity is
suppressed compared to ${\rm R_{AA}^{HF}}$ (Cu+Cu) at mid-rapidity,
suggesting the possibility that cold nuclear matter effects are larger
at forward rapidity. The observed suppression is consistent with a
calculation~\cite{Sharma} that includes the effects of heavy-quark
energy loss and in-medium heavy-meson dissociation, as well as cold
nuclear matter effects due to shadowing and initial state energy loss
due to multiple scattering of incoming partons before they interact to
form the $c\bar{c}$ pair.

\begin{figure}[h]                                                                                    
\begin{center}
\begin{minipage}{18pc}                                                                               
\includegraphics[width=18pc]{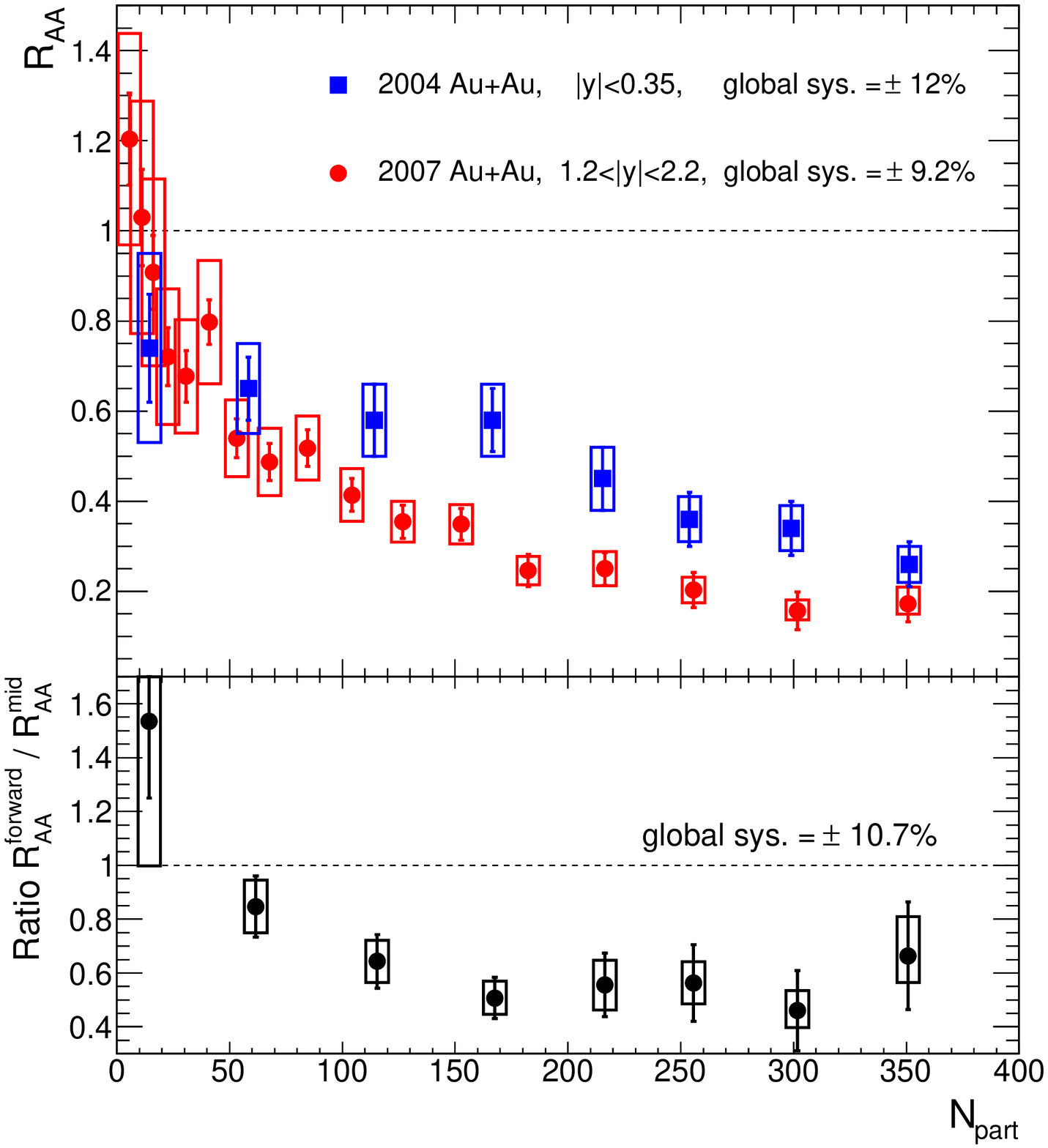}
\caption{\small \label{fig7} J/$\psi$ ${\rm R_{AA}^{HF}}$ as a
  function of $N_{part}$ in Au+Au collisions at 200\,GeV. Error bars
  represent the statistical uncertainties. The boxes represent the
  point-to-point correlated systematics. The lower panel contains the
  ratio of forward rapidity to mid-rapidity~\cite{PHE11A}.}
\end{minipage} 
\hspace{0.5pc}%
\begin{minipage}{18pc}                                                                               
\includegraphics[width=18pc]{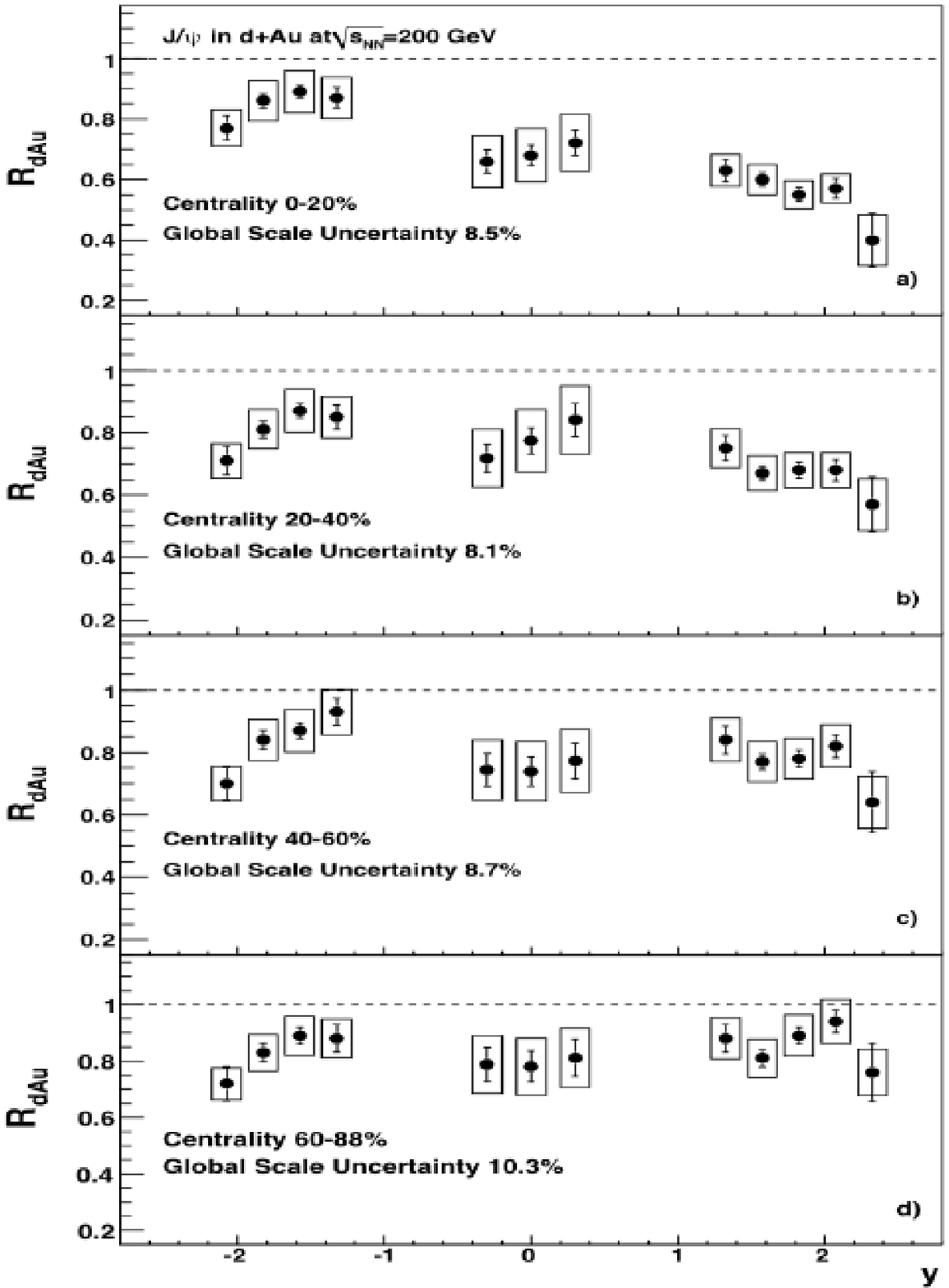}
\caption{\small \label{fig8} J/$\psi$ ${\rm R_{dAu}^{HF}}$  versus rapidity for four
  centrality bins in d+Au collisions at 200\,GeV~\cite{PHE11B}.}
\end{minipage}                                                                                       
\end{center}
\end{figure}        

\section{Heavy quarkonia}

Heavy quarkonia have long been proposed as a sensitive probe of the
color screening length and deconfinement in the quark-gluon
plasma~\cite{Mue05}. The picture that was originally proposed is
complicated by competing effects, which modify quarkonia
production and survival in cold and hot nuclear matter. Heavy
quarkonia measurements are made in PHENIX either by detecting opposite
sign electrons at mid-rapidity ($\mid$y$\mid <$0.35) or by detecting
opposite sign muons at forward rapidity (1.2~$<\mid$y$\mid<$~2.2),
reconstructing the invariant mass of the di-lepton pair, and
subtracting the continuum background~\cite{PHE07}. The nuclear
modification factor, ${\rm R_{AA}}$, for J/$\psi$ as a function of
centrality ($N_{part}$) at mid-rapidity and forward rapidity from Au+Au
collisions is shown in Fig.~\ref{fig7}. The data show that the
suppression of J/$\psi$ at forward rapidity is stronger than at
mid-rapidity. The comparison of the experimental data to the most
recent theoretical calculations that incorporate a variety of physics
mechanisms including gluon saturation, gluon shadowing, initial-state
parton energy loss, cold nuclear matter breakup, color screening, and
charm recombination has been published by PHENIX in
Ref.~\cite{PHE11A}. Figure~\ref{fig8} shows the J/$\psi$ ${\rm
  R_{dAu}^{HF}}$ as a function of rapidity for each of the four
centrality bins. The forward rapidity data for the most central
collisions show a suppression of about 50\%. Using these measurements,
various models have been suggested to describe the cold nuclear matter
effects on J/$\psi$ production used by PHENIX in Ref.~\cite{PHE11B}.

\begin{figure}[!]                                                                                    
\begin{center}
\includegraphics[height=30pc]{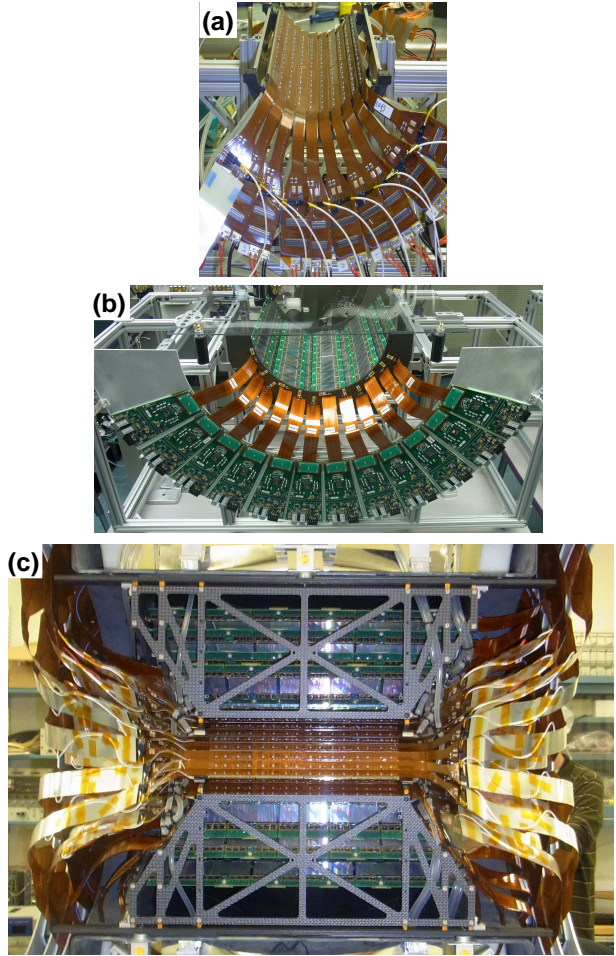}
\caption{\small \label{fig9} The PHENIX Silicon Vertex Tracker
  (VTX). Panels (a), (b) are end-views and (c) is a side view. Panels
  (a) and (b) show the ladders of half barrels of pixel and stripixel
  detectors, respectively. Panel (c) shows assembled half
  VTX~\cite{RNVTX,RNNews}.}
\end{center}
\end{figure}        

\begin{figure}[!]
\begin{center}                                                                                    
\begin{minipage}{18pc}                                                                               
\includegraphics[width=20pc]{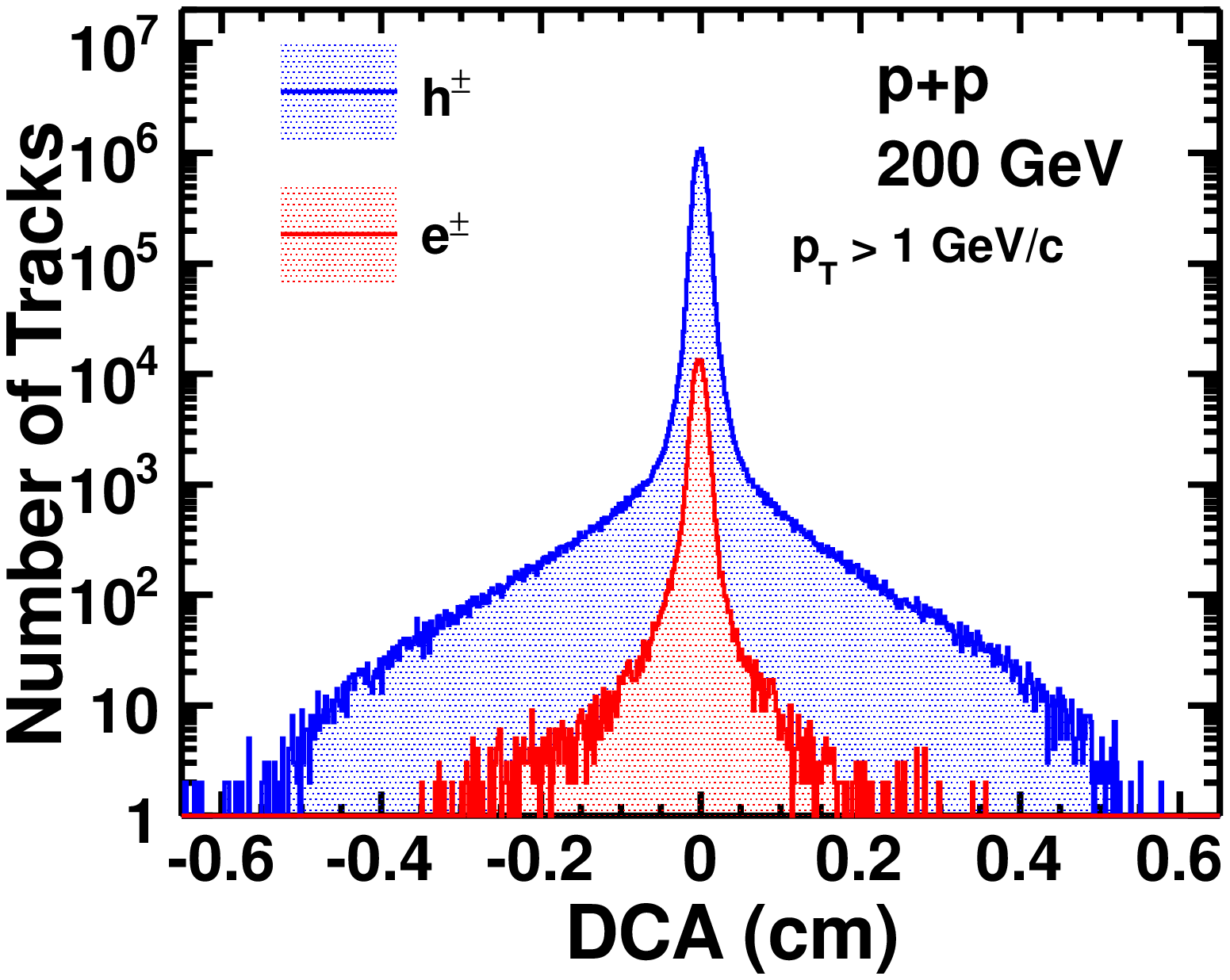}
\caption{\small \label{fig10} DCA distribution for charged hadrons
  and electrons (candidates from HF decays) in $p+p$ collisions at 200\,GeV.
  The DCA are integrated for transverse momentum $p_T$$>$1\,GeV/$c$.}
\end{minipage} 
\hspace{0.5pc}%
\begin{minipage}{18pc}                                                                               
\includegraphics[width=20pc]{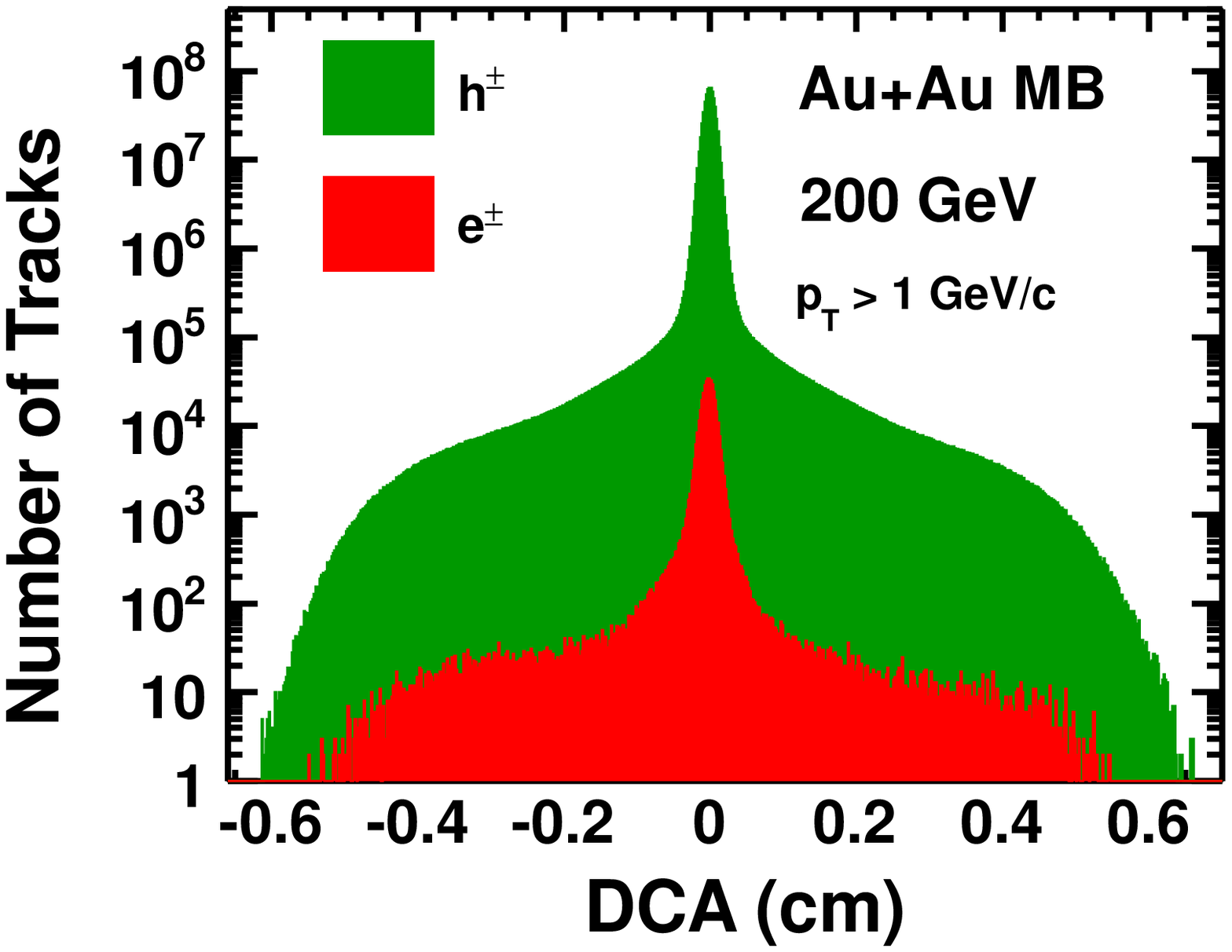}
\caption{\small \label{fig11} DCA distribution for charged hadrons
  and electrons (candidates from HF decays) in minimum-bias Au+Au
  collisions at 200\,GeV. The DCA are integrated for transverse momentum
  range $p_T$$>$\,1\,GeV/$c$.}
\end{minipage}
\end{center}                                                                                       
\end{figure}       

\section{New era of heavy flavor measurements : PHENIX Silicon Vertex Tracker}

For these early results, PHENIX was not able to distinguish electrons
coming from the semi-leptonic decay of $D$~and~$B$ mesons
independently. In order to understand these medium effects in more
detail it is imperative to {\em directly} measure the nuclear
modification and the flow of $D$~and~$B$ mesons separately. Based on
this motivation and impelled by the exciting physics we had already
uncovered, in December 2010, the PHENIX Collaboration opened new era for
measuring heavy flavor at RHIC by installing new detector called the
Silicon Vertex Tracker (VTX)~\cite{RNVTX,RNNews}. The VTX was
commissioned with $p+p$ at 500\,GeV, and took data of $p+p$ and
Au+Au during Run-11 and, $p+p$, Cu+Au and U+U during Run-12 at
RHIC. The VTX detector consists of four layers of barrel detectors
located in the region of pseudorapidity $\mid \eta \mid <$~1.2 and
covers almost the full azimuth.  The two inner barrels of the VTX detector
consist of silicon pixel detector.  and the two outer barrels of the
VTX are constructed using silicon stripixel sensors. Figure~\ref{fig9}
shows the ladders of half barrels of pixel and stripixel, as well one
complete half of the assembled VTX detector.

The key element needed in directly identifying $D$ and $B$ mesons in
PHENIX is the ability to measure the collision vertex position with
high accuracy and to compare that with the trajectory for each
track. $D$ and $B$ mesons decay into light mesons or leptons before
reaching the detectors, thus the daughter particles are observed. As
these do not originate at the collision vertex position, a large
distance (the decay-length of $D^0$ and $B^0$ are 123 and 457~$\mu$m,
respectively) occurs between the vertex position and the daughter
particle trajectory. Based on the VTX detector measurements of the
primary vertex, the beam size, and individual tracks for each
collision like in Au+Au or in $p+p$ collisions, the reconstructed
distance of closest approach (DCA) distribution from primary vertex
has been determined. In the present results from $p+p$ collisions at
200\,GeV, the DCA distribution was obtained using beam center (instead
primary vertex) to avoid auto-correlation (in $p+p$ there are only few
tracks). DCA distributions in $p+p$ and Au+Au collisions at 200\,GeV
obtained from RHIC Run-11 and Run-12 are presented in Fig.~\ref{fig10}
and Fig.~\ref{fig11}, respectively. Figure~\ref{fig11} shows that the
DCA resolution of individual tracks to the primary vertex,
$\sigma$(DCA), is about 70 $\mu$m. The achieved DCA resolution is
sufficient to distinguish $D$- and $B$-mesons from the difference of
their decay-length. These measurements using PHENIX VTX detector
clearly illustrate that this detector is working as expected and
future heavy flavor physics program in PHENIX is very promising.

\end{document}